\documentclass[conference]{IEEEtran}
\IEEEoverridecommandlockouts

\usepackage{cite}
\usepackage{amsmath,amssymb,amsfonts}
\usepackage{algorithmic}
\usepackage{graphicx}
\usepackage{textcomp}
\usepackage{xcolor}
\usepackage[colorlinks=false, pdfborder={0 0 0}, breaklinks]{hyperref}
\usepackage[inline]{enumitem}
\usepackage{makecell}
\usepackage{soul}
\usepackage{booktabs}
\usepackage{multirow}
\usepackage{multicol}
\usepackage{enumitem}
\usepackage{caption}
\usepackage{subcaption}
\def\BibTeX{{\rm B\kern-.05em{\sc i\kern-.025em b}\kern-.08em
    T\kern-.1667em\lower.7ex\hbox{E}\kern-.125emX}}

\begin{document}
\bstctlcite{IEEEexample:BSTcontrol}

\title{Content Adaptive Encoding For\\ Interactive Game Streaming

}

\author{Shakarim Soltanayev$^*$, Odysseas Zisimopoulos$^*$, Mohammad Ashraful Anam, Man Cheung Kung,\\ Angeliki Katsenou, and Yiannis Andreopoulos\\
Sony Interactive Entertainment \thanks{* Equal contribution, the author listing is random. We would like to thank Ilya Fadeev for his contributions.}}

\maketitle

\begin{abstract}
Video-on-demand streaming has benefitted from \textit{content-adaptive encoding} (CAE), i.e., adaptation of resolution and/or quantization parameters for each scene based on convex hull optimization. However, CAE is very challenging to develop and deploy for interactive game streaming (IGS). Commercial IGS services impose ultra-low latency encoding with no lookahead or buffering, and have extremely tight compute constraints for any CAE algorithm execution. We propose the first CAE approach for resolution adaptation in IGS based on compact encoding metadata from past frames. Specifically, we train a convolutional neural network (CNN) to infer the  best resolution from the options available for the upcoming scene based on a running window of aggregated coding block statistics from the current scene. By deploying the trained CNN within a practical IGS setup based on HEVC encoding, our proposal: \textit{(i)} improves over the default fixed-resolution ladder of HEVC by 2.3 Bjøntegaard Delta-VMAF points; \textit{(ii)} infers using 1ms of a single CPU core per scene, thereby having no latency overhead. 
\end{abstract}


\section{Introduction}


Interactive game streaming (IGS) is one of the most demanding use cases of networked media applications, pushing the limits of video encoding technologies due to its extremely stringent requirements on end-to-end latency, high visual fidelity, and the very dynamic nature of game content~\cite{BarmanIEEETCSVT2022}. 
Unlike traditional video-on-demand (VoD) services, where the use of buffering can ensure a smooth playback experience with high coding efficiency, IGS requires real-time encoding and delivery to support the interaction between the user controls and the cloud server running the game. For example, when playing 60 frame-per-second (fps) games, video encoding must be performed within 16ms without buffering ~\cite{chen2023gamival}. Developing content-adaptive encoding (CAE) methods for IGS is very challenging because:
\begin{itemize}
\item CAE essentially exploits the ability to preprocess and buffer content under relaxed latency constraints. 
\item CAE algorithms require significant CPU resources and usually carry out multiple full or partial encodes per scene to gather enough encoding statistics and make scene-adaptive decisions. 
\end{itemize}
This shows that existing CAE approaches cannot be adapted for IGS services, i.e., we must rethink how to introduce content statistics for per-scene adaptivity within IGS. 

\subsection{Related Work}
\label{ssec: Related}

In terms of algorithms,  CAE approaches essentially boil down to the static or dynamic construction of a bitrate ladder for each scene to be encoded. Static ladder construction methods often rely on fixed content-agnostic~\cite{HLS_ladder_ref} or heuristic-based approaches~\cite{ReznikPV2018, bourtsoulatze2019deep, wu2020fast, paul2022efficient, TeliliACMTransMCCA2025}, primarily focused on rate-quality (RQ) curves that are either estimated or derived experimentally. 
Dynamic ladder optimization techniques, such as per-title/per-scene encoding ladders~\cite{netflix_paper, bourtsoulatze2019deep, wu2020fast, paul2022efficient, HAS_survey, KokaramICIP2018}, perceptual bitrate allocation \cite{jingwen_vmaf_jnd_sur, Menon_ICME2022}, and pre-trained machine learning-based quality enhancement~\cite{Katsenou_IEEEOJSP2021, bourtsoulatze2019deep, ensemble_learning_vvc_ladder, Menon_ICME2022, menon2024convexhull_xpsnr, Krishna_PCS2024}, are commonly employed to balance bitrate and visual quality.
All these approaches rely on compute-intensive models (e.g., using pixel-based feature extraction and recurrent learning), tend to apply more than one encoding per scene, and assume that (partial) buffering of the next scene is possible. This makes them impractical for IGS where no buffering or significant CPU resources are available, and only a single encoding can take place per frame.

Recently, research focus has shifted to low-latency streaming ~\cite{tashtarian2024artemis, GhasempourJETCAS2025}.  All current efforts are on online broadcasting use cases, where certain amount of latency tolerance is possible, e.g., in the order of 1s (15 to 60 frames). Therefore, all these techniques are based on variations of server-side optimization of temporal prediction structures, temporal quality smoothing, and encoder lookahead, which do not apply to IGS, due to its ultra-low latency requirement. Overall, the constraints of IGS are: \textit{(i)} for each scene, users always prefer the highest resolution possible under their connection bitrate ~\cite{BarmanIEEETCSVT2022}; \textit{(ii)} the reactive nature of IGS imposes no encoder lookahead and strict video buffer and latency limits; \textit{(iii)} the combination of the first two constraints can lead to frame drops at the sender side, which can be experienced as ``game stuttering'' by the user and must be avoided; \textit{(iv)} there is very limited compute available to infer the optimum CAE setting per scene, as all IGS deployments push the use of the CPU, GPU and ASIC encoding resources to their limits. Since none of the existing CAE proposals meets these four conditions, there is a need for specialised CAE strategies for IGS. 

\begin{figure}
    \centering
    \includegraphics[width=\linewidth]{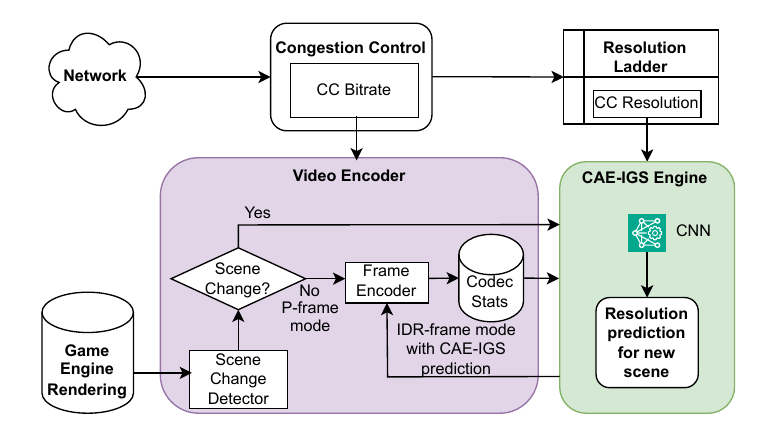}
    \caption{Diagrammatic overview of the CAE-IGS framework.}
    \label{fig: method}
    \vspace{-1em}
\end{figure}

\subsection{Novelty and Contribution}
\label{ssec: Contribution}
We propose CAE for IGS (CAE-IGS), based on a learnable per-scene resolution adaptation approach via a light-weight convolutional neural network (CNN) model that runs on CPU. Our proposal makes the following contributions: 
\begin{itemize}
\item CNN training to infer the best resolution for the upcoming scene based on the optimal resolution per bitrate (selected offline from the available ones). 
\item Instead of utilizing video frames as input (which is computationally infeasible in IGS), the proposed CNN infers the resolution to use for the next scene by ingesting macroblock line-aggregated statistics, i.e., coding tree block (CTB) stats for High Efficiency Video Codec (HEVC), that are logged from the past $N$ frames by the encoder without additional overhead (Fig. \ref{fig: method}). 
\item Resolution adaptation takes place at scene changes (Fig. \ref{fig: method}), where an IDR (instantaneous decoder refresh) frame is \textit{de-facto} used, thereby avoiding any extra IDRs that could cause adverse effects in RQ or congestion control. 
\item Unlike all existing CAE approaches, the proposed CAE-IGS \textit{does not}: \textit{(i)} carry out any full or partial frame encodings to get RQ estimates, \textit{(ii)} use any lookahead frames or lookahead statistics, \textit{(iii)} incur any notable compute overhead, as its inference time is within 1ms on a single CPU core, thereby imposing no latency overhead.
\end{itemize}

CAE-IGS is evaluated with challenging top-tier 1080p-60fps gaming content and HEVC encoding. Our test conditions incorporate the occurrence of frame drops due to the strict congestion control conditions of ultra-low latency game streaming. This scenario corresponds to IGS for portable devices, such as the Sony PlayStation Portal Remote Player, which experience a range of RQ conditions when users move to different environments. 
Under the same bitrate targets, the results demonstrate 2.3-point improvement in VMAF versus the static ladder method, without detrimenting (and even slightly improving) frame drop statistics. This improvement is predominantly in the ``active'' quality region of VMAF (between 50-80).   

The remainder of the paper is organized as follows. Section \ref{sec: Proposed} describes the operation of CAE-IGS, Section \ref{sec: setup} presents the experimental setup, and Section \ref{sec: results} discusses our results. Finally, Section \ref{sec: connclusion} concludes the paper. 



\section{Proposed CAE-IGS Method}
\label{sec: Proposed}
The  operation of our proposal is illustrated in Fig.~\ref{fig: method}. Within the overall IGS system, the network conditions cause the congestion control to set a connection (maximum) bitrate (CC bitrate). This bitrate is used by the  resolution ladder of the  encoder to derive the default resolution to use for that bitrate (CC resolution). For each point in the ladder, the proposed CAE-IGS engine is trained to infer the spatial resolution for the next scene using codec statistics from $N$ past frames\footnote{In our experiments, $N$ is set to 60. This was experimentally found to strike a good balance between input codec stats' predictive power and inference complexity.}. The CAE-IGS prediction may differ from the default CC resolution.  

As shown in Fig.~\ref{fig: method}, when a new frame comes from the game engine, the scene change detector decides if a scene change is to be triggered. If so, the scene change detector triggers the CAE-IGS resolution prediction to be applied in the coming scene and mandates the frame encoder to encode the new frame as an IDR frame. Otherwise, the new frame is encoded as a predicted (P) frame with the current resolution setting. In this way, resolution adaptation is only carried out during scene changes, which are triggered purely based on the game activity. 

Since the CNN inference uses only past frames for the next-scene resolution prediction, this will occasionally lead to suboptimal predictions in comparison to inferring for the current scene \textit{a-posteriori}. However, this is taken into account during the offline training of the CAE-IGS CNN, as our CNN model is trained to infer the best resolution of the \textit{next scene}. In addition, to ensure that the occasional mispredictions are not visually detrimental, we only allow for 1 step adaptation of the CC resolution ladder point, e.g., from 540p to 720p (and similar for other points, as shown in Table \ref{tab: CronosLadder}). This sacrifices some potential RQ gains, but forms a safer option for practical deployment.

\subsection{Exploring the Rate-Quality Parameter Space}
\label{ssec: RQspace}
To train our CAE-IGS model, we first explore the RQ parameter space across spatial resolutions. We perform grid encodings across resolutions $S$ for a wide range of target bitrate $R$ values that cover the operational CC bitrate values. In this way, we construct the optimal convex hull per video scene $v$. From this convex hull, $C$, we construct the optimal rate-resolution ladder $L_v = \{R_v, S_v \}$ that is encapsulating the RQ parameter space. These ground-truth optimal grid encoding points are termed as the \textit{Optimal Ladder}.

\begin{figure}
    \centering
    \includegraphics[width=.86\linewidth]{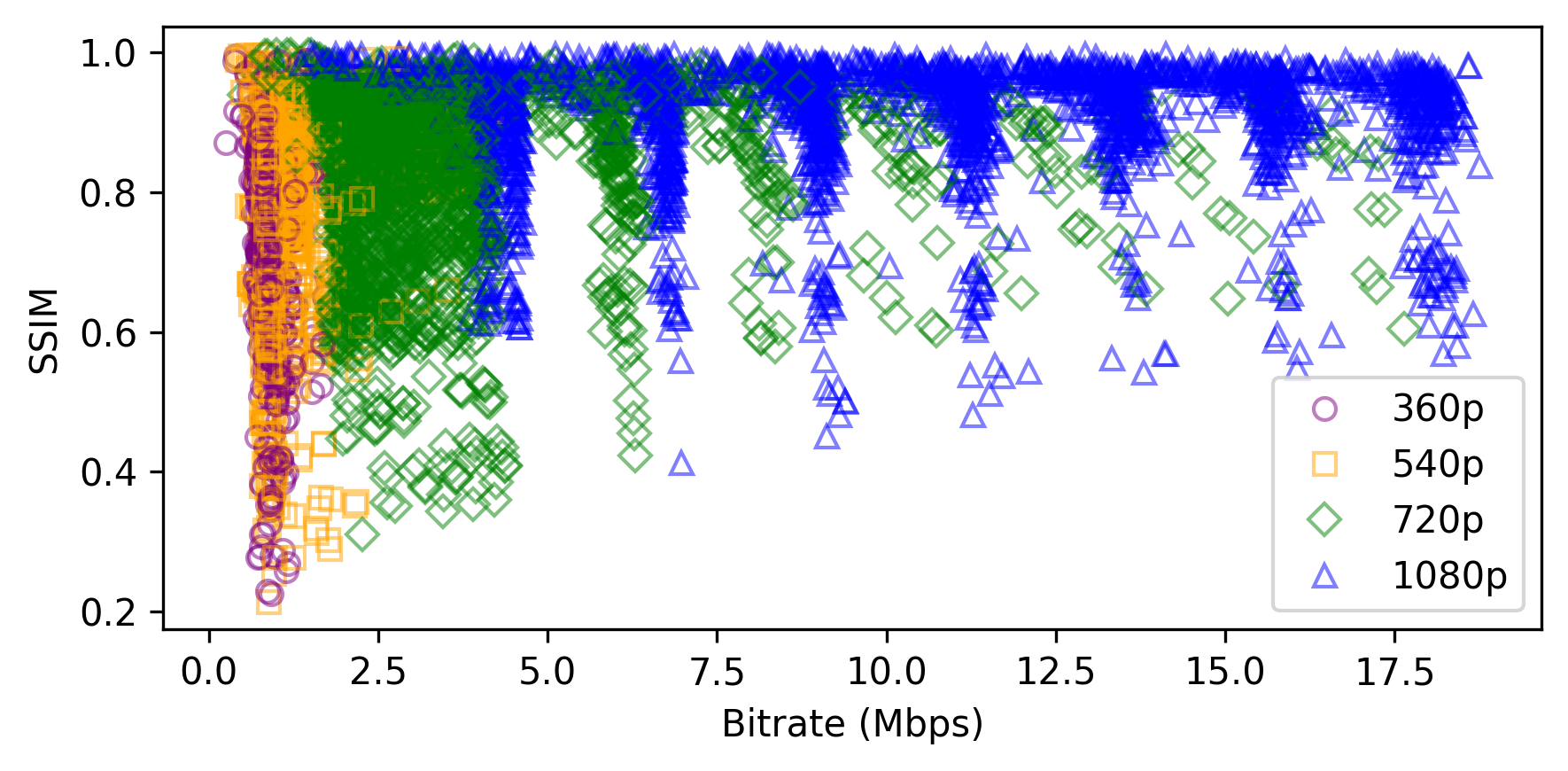}
    \vspace{-0.6em}
    \caption{Convex hull composition in RQ space. }
      \vspace{-1em}
    \label{fig: ParamSpace}
\end{figure}

Figure~\ref{fig: ParamSpace} illustrates an example of the RQ space across different spatial
resolutions, i.e. $\mathbb{S} = \{360\text{p}, 540\text{p}, 720\text{p}, 1080\text{p}\}$, for 500 game scenes of 1080p native resolution (60 fps, 8 bits depth, with 420 colour sampling). Naturally, higher resolutions and higher bitrate values result in better video quality (in this case expressed in structural similarity index measure (SSIM)~\cite{Wang_SSIM} values). However, we see from the figure that the resolution switching points on the convex hulls are not the same for all sequences. They depend on the compression performance on different sequences. For example, game scenes that contain dynamic textures and non-linear motion are usually harder to compress~\cite{katsenou2022study}.

\subsection{Selected Features for the CAE-IGS Engine}
\begin{figure}
    \centering
    \includegraphics[width=\linewidth]{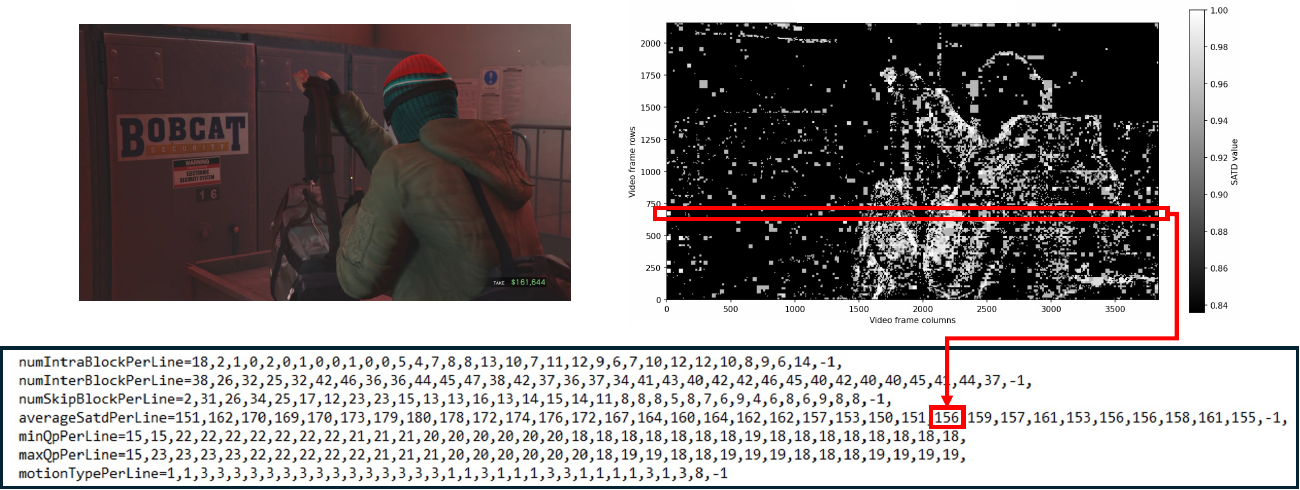}
    \vspace{-1.2em}
    \caption{Example of codec statistics: a 4K video frame is depicted along its SATD frame (averages of residual errors for each prediction block area) and the per-line stats. Every statistic comprises 34 values, as this is a 2160p frame that has 34 rows of HEVC CTBs.} 
     \vspace{-1.7em}
    \label{fig:PerLineFeatures}
\end{figure}

The video encoder produces per-frame statistics (metadata) after encoding a video clip. We parse a selection of these statistics as inputs to our models. This subset of features selected was based on inspection of their correlation to compression performance and also highlighted by past research~\cite{AfonsoPCS2016, katsenou2022study}. As shown in Fig.~\ref{fig:PerLineFeatures}, the features are per-line encoding statistics, i.e., the averages of a row of macroblocks (or CTBs in HEVC). We use the prediction mode statistics (\texttt{numIntraBlockPerLine}, \texttt{numInterBlockPerLine}, \texttt{numSkipBlockPerLine}), an expression of the sum of absolute pixel differences (SATD - \texttt{averageSatdPerLine}), the quantisation parameter (QP) statistics (\texttt{minQpPerLine}, \texttt{maxQpPerLine}), and a statistic related to the type of motion (\texttt{motionTypePerLine}). Since the length of per line stats depends on the resolution of the video, we reshape them to a common size using bilinear interpolation.

\begin{figure*}[!th]
    \centering
    \includegraphics[width=\linewidth]{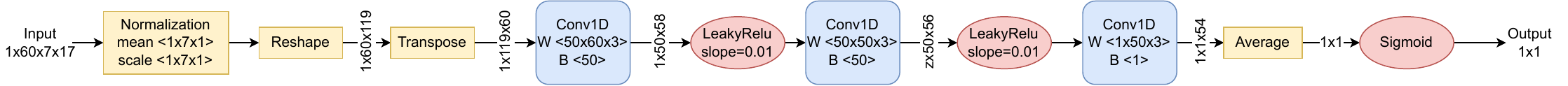}
    
    \caption{Proposed CNN model architecture. For each bitrate, the input comprises the 7 per-line stats (Fig.~\ref{fig:PerLineFeatures}) of the previous $N=60$ frames (with each stat rescaled to 17 values). The output  infers which of the two CAE-IGS resolutions of Table \ref{tab: CronosLadder} should be used depending on the bitrate zone.}

    \vspace{-1em}
    \label{fig:Network}
\end{figure*}

\subsection{CNN Model Architecture}
\label{ssec: NetwArchitecture}
As explained in Section~\ref{ssec: RQspace}, from the exploration of the parameter space we gain insights about the optimal selection of resolutions across bitrates. In order to design our network we take into account the resolution distribution across the zones as well a condition of resolution monotonicity so that we ensure smooth transitions across the ladder. Based on the resolution distribution, we further divide the bitrate range into distinct zones (see Table~\ref{tab: CronosLadder}) and solve the resolution allocation as a binary classification problem within each zone. Thus, we train a separate binary classifier for each zone to predict the optimal resolution between the two candidate choices of Table~\ref{tab: CronosLadder}. To generate ground‐truth labels, we use the Optimal Ladder $L_v = \{R_v, S_v \}$.  
Then for a given target bitrate, we identify the hull points with the closest target bitrates and select the point with the highest quality metric. The resolution associated with this optimal point is treated as the ground truth. We then map that point’s resolution to a binary label assigning ``0'' to the lower‐resolution candidate and ``1'' to the higher. For example, for target bitrates in zone 2, the choice is between 720p and 1080p, so ``0'' would mean 720p and ``1'' would mean 1080p.

Fig.~\ref{fig:Network} depicts the CAE-IGS CNN architecture. The input to our model is a four-dimensional array of shape (1x60x7x17), where 60 represents the number of frames and 7 denotes the number of per-line statistical features. The final dimension, 17, corresponds to the length of the per-line statistics for 1080p resolution encoding; for lower resolutions, this dimension is smaller and is upsampled to 17 using bilinear interpolation. Each of the 7 per-line statistics is standardized individually using the global dataset mean and standard deviation. The last two dimensions are then flattened to produce an input suitable for one-dimensional CNN (Conv1D). 
The network architecture comprises of three Conv1D layers with LeakyReLU activations, culminating in a final Conv1D layer with a single output channel. This output is passed through global average pooling and a sigmoid activation to produce the final prediction. The model is trained using binary cross-entropy loss between the prediction and the ground truth label. During inference, the threshold of 0.5 is applied to the output to determine the optimal resolution in an unbiased manner. Training of each of the four CNN models (one per zone of Table~\ref{tab: CronosLadder}) is performed on an NVIDIA A100 GPU with a batch size of 64 over 50,000 iterations. The  CNN architecture of Fig.~\ref{fig:Network} is hand-optimized with custom C++ code suitable for CPU deployment. Inference on 60-frame input stats runs in approximately 1ms on a single AMD Zen 2 CPU core (RDNA 2 architecture) by invoking the appropriate CNN model for each zone of Table~\ref{tab: CronosLadder}.

\section{Experimental Set-up}
\label{sec: setup}
This section describes the experimental setup for the training and evaluation of the CAE-IGS solution.

\subsection{Compared Methods}
Since all external CAE ladder construction methods cannot operate on the IGS environment of our proposal due to their latency and CPU runtime requirements, we have considered and tested the methods described below.
\begin{enumerate}
    \item \textit{Optimal Ladder}: This corresponds to brute-force selection of the best resolution from the CAE-IGS options of the fourth column of Table \ref{tab: CronosLadder}. That is, every scene is \textit{a-posteriori} encoded as a batch, metrics are measured on both options and then the best resolution is selected. It serves as an upper bound that is generally unachievable. 
    \item \textit{Static Ladder}: This is the static bitrate-resolution ladder of the underlying HEVC encoder, shown in the third column of Table \ref{tab: CronosLadder}. This static ladder was derived via extensive offline experimentation with aggregated statistics across multiple bitrate zones and was found to optimize the aggregate RQ results over a range of gaming scenes, while keeping frame drops at a minimum level in order to ensure smooth gameplay experience.
\end{enumerate}

\subsection{Dataset Description}
The dataset used for training and testing comprises a total of 120 gameplay video clips drawn from 38 unique game titles from the PlayStation Plus Premium service. These titles span a broad range of genres and production scales, encompassing high-budget AAA, AA, and indie games. 
All clips from AAA titles in the dataset exhibit photorealistic graphics, high realism, and complex gameplay physics effects, often featuring large open worlds, and cinematic storytelling. In contrast, the indie and AA games typically present stylised visuals, abstract or simplified realism, and focused gameplay loops that emphasise creativity, narrative, or arcade-style (platform games) interactions. The clips also reflect a diversity of content, including fast action, menus, cutscenes, and special effects such as flashing segments, floating particles, dynamic textures, reflections, and film grain, thereby offering a varied and representative sample of modern gaming content. All sequences were captured natively at 2160p, 60fps, 8bits depth, and 420 color sampling. For the evaluation we split the dataset into 80\% training and 20\% testing. Finally, each of the video clips has a varying number of scene changes, usually between two to six.

\subsection{Encoding Configuration}
\label{sec: setup-static}
The sequences were downscaled to multiple spatial resolutions, i.e., \{1080p, 720p, 540p, 360p\} with the lanczos-2 filter. The maximum streamed resolution is 1080p, since we focus on the challenging case of gaming on portable device at the bitrates of Table \ref{tab: CronosLadder}. For all experiments, we used HEVC/H.265~\cite{HEVC} with an ultra-low delay recipe and no lookahead encoding or buffering. Beyond HEVC, CAE-IGS can be extended to other codecs with similar stats to the ones of Fig. \ref{fig:PerLineFeatures}, but this is outside the scope of this paper.

\begin{table}[h]
\caption{The combinations of spatial resolution and bitrate ranges of the Static Ladder and the CAE-IGS ladder.}
\vspace{-1em}
\begin{center}
\begin{tabular}{crrr}
\toprule
\textbf{Zone} & \textbf{Bitrate (Mbps)}& \textbf{Static Resolution} & \textbf{CAE-IGS Resolution}\\
\midrule
0 & 1-2& 360p & 360p or 540p\\ 
\hline
1 & 2-5& 540p & 540p or 720p\\
\hline
2 & 5-10& 720p & 720p or 1080p\\
\hline
3 & 10-20& 1080p & 720p or 1080p\\
\bottomrule
\end{tabular}
\vspace{-1em}
\label{tab: CronosLadder}
\end{center}
\end{table}

\section{Evaluation and Results}
\label{sec: results}
For the evaluation of the proposed solution we selected five target bitrate points from the zones in Table~\ref{tab: CronosLadder}, specifically \{2, 3.5, 7.5, 12.5, 17.5\} Mbps, to encode the sequences in the test set using the Static Ladder, the Optimal Ladder and CAE-IGS. We measured the Bjonteegard Delta-rate (BD-Rate) and BD on quality metrics on three widely used metrics in video quality assessment, namely VMAF, SSIM-Yb, and PSNR-Y. For these calculations we used the libvmaf library \cite{VMAF}. Since all BD  calculations require RQ convexity, all ladder points that lead to non-convex hulls are dropped from our measurements.

\begin{figure}[t]
    \captionsetup[subfigure]{font=footnotesize}
     \centering
     \begin{subfigure}[b]{0.49\columnwidth}
         \centering
         \includegraphics[width=\textwidth]{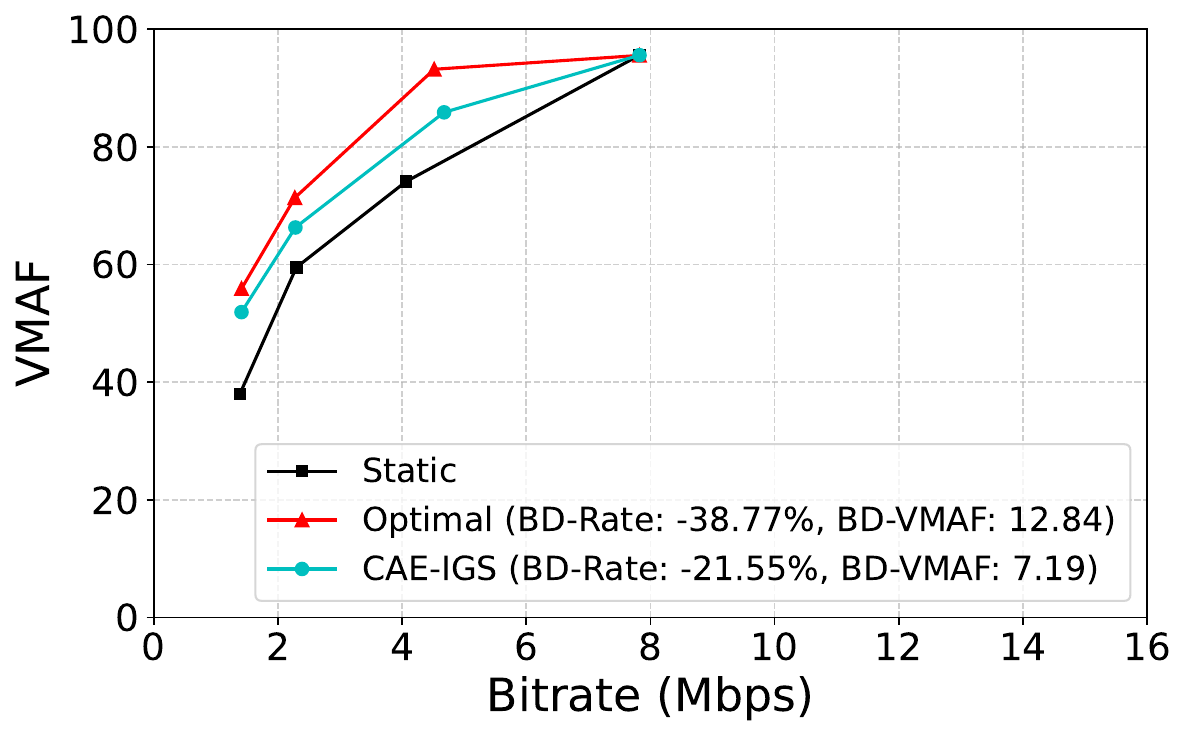}
         \caption{Test Sequence 1}
         \label{fig: Spider}
     \end{subfigure}
     \hfill
     \begin{subfigure}[b]{0.49\columnwidth}
         \centering
         \includegraphics[width=\textwidth]{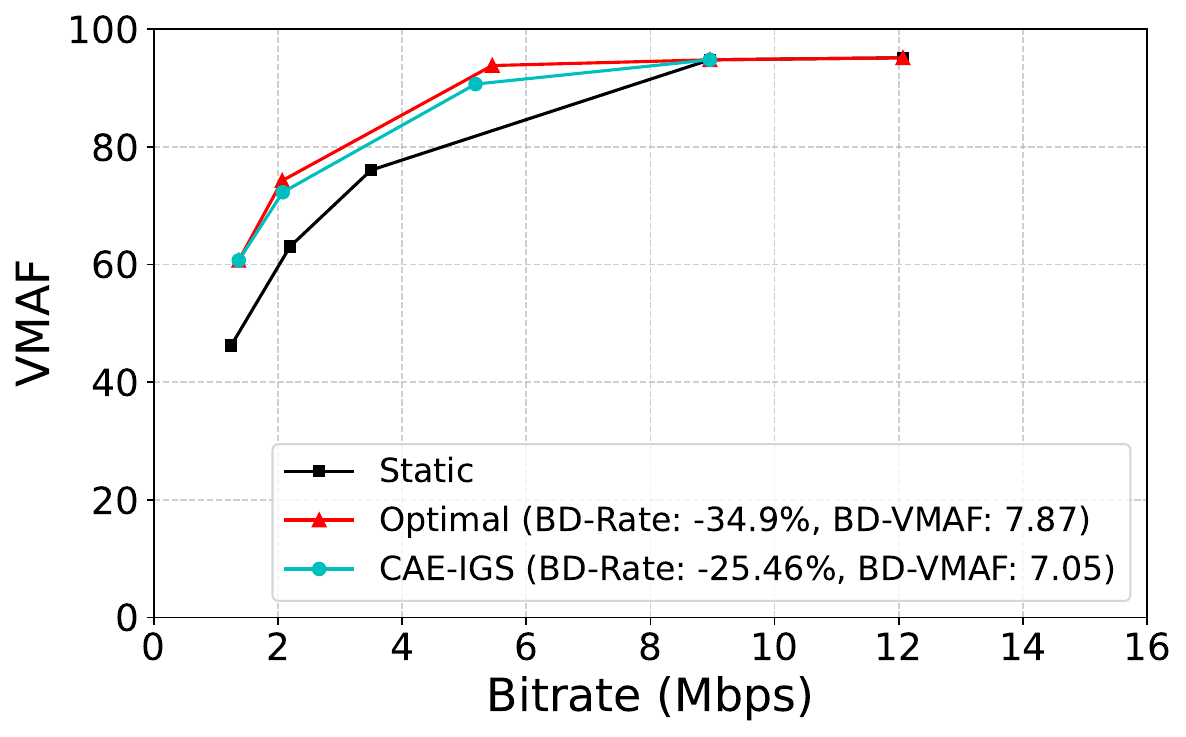}
         \caption{Test Sequence 2}
         \label{fig: Spider}
     \end{subfigure}

          \begin{subfigure}[b]{0.49\columnwidth}
         \centering
         \includegraphics[width=\textwidth]{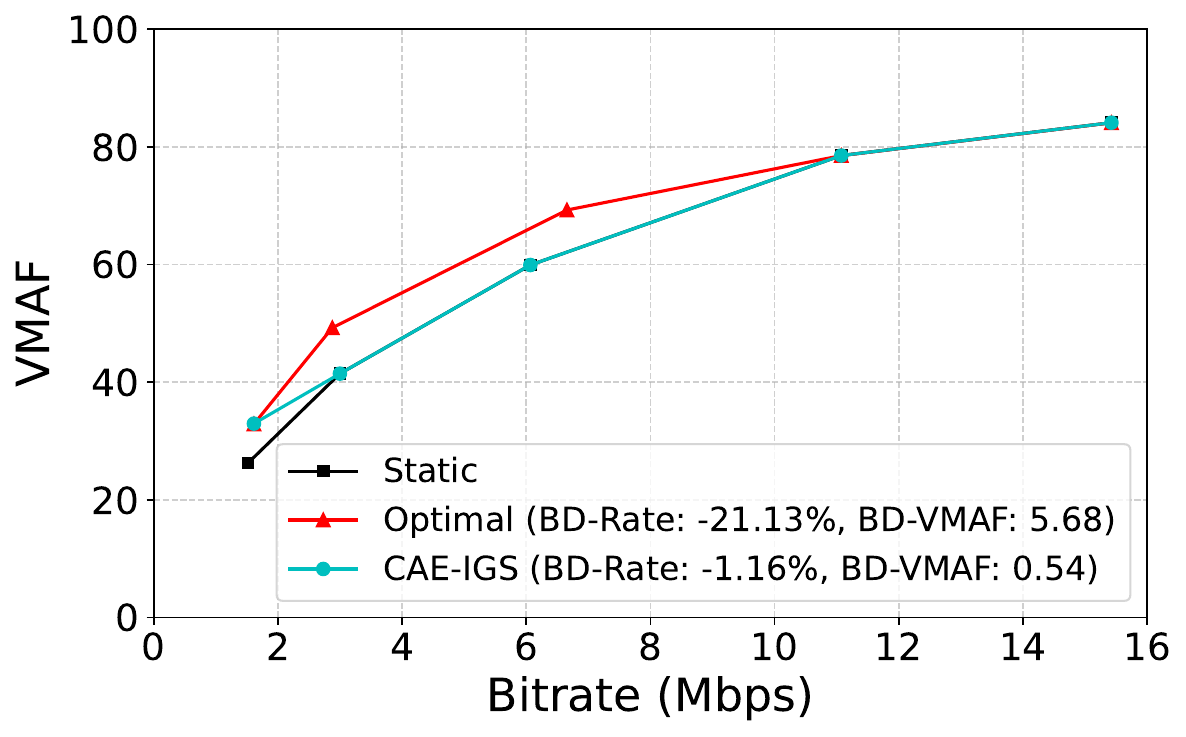}
         \caption{Test Sequence 3}
         \label{fig: Devil}
     \end{subfigure}
     \begin{subfigure}[b]{0.49\columnwidth}
         \centering
         \includegraphics[width=\textwidth]{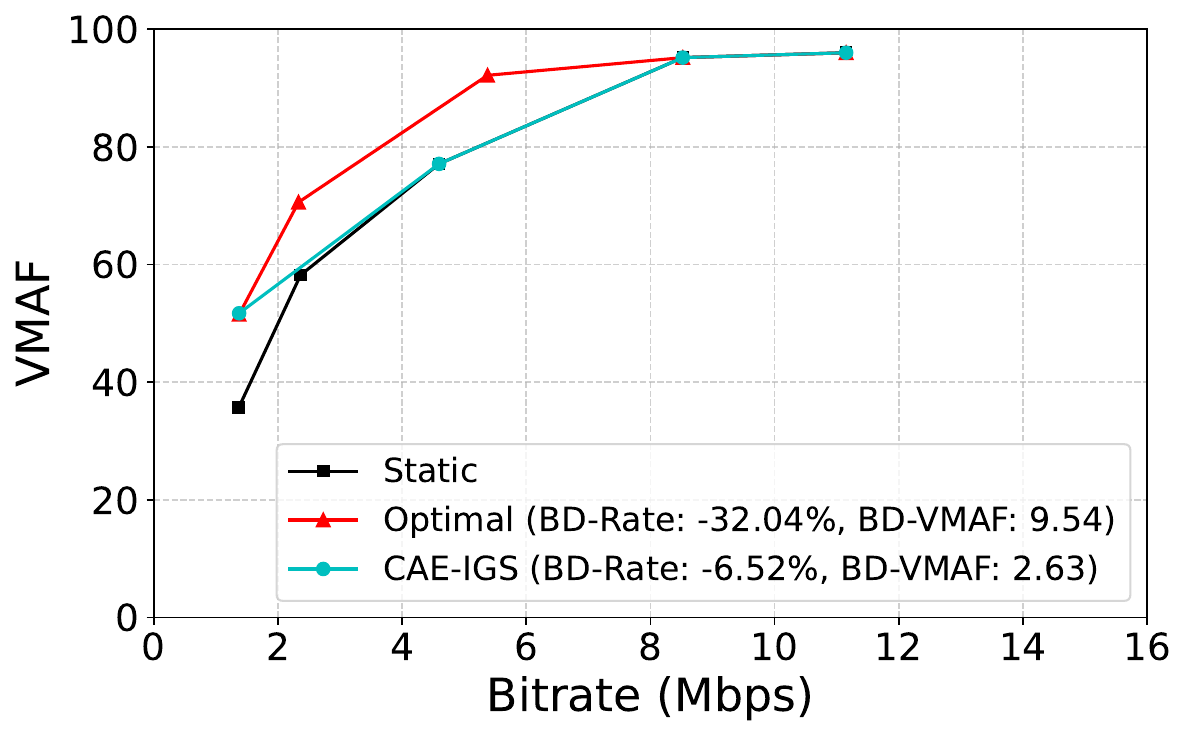}
         \caption{Test Sequence 4}
         \label{fig: FF7}
     \end{subfigure}
     
        \caption{Examples of resulting ladders for four different test sequences that demonstrate most common improvement patterns of the proposed CAE-IGS and Optimal Ladder over the baseline Static Ladder solution.}
        \vspace{-1em}
        \label{fig: ExamplesLadders}
\end{figure}

Representative examples of Rate-VMAF plots are shown in Fig.~\ref{fig: ExamplesLadders}. For the first two sequences significant bitrate savings and VMAF increase are achieved across most target bitrates. Furthermore, for Test Sequence 2, CAE-IGS operates very closely to the Optimal Ladder. For the sequences 3 and 4, the gains are in the lower bitrate zones, which are the main concern in streaming in terms of quality improvement.

\begin{table}[ht]
\centering
\caption{BD metrics and $\Delta$Framedrops (\%) against the Static Ladder of Table \ref{tab: CronosLadder}. Optimal Ladder is presented as an upper bound of performance, as it is not practical for IGS deployments. The arrows indicate preferable trends.}
\label{tab:bd_metrics}
\begin{tabular}{l|r|r}
\toprule
\textbf{Metric} & \textbf{Optimal Ladder}  & \textbf{Proposed CAE-IGS} \\
\midrule
BD\textsubscript{V}-Rate (\%) $\downarrow$ & -21.0 & -7.0 \\
BD-VMAF $\uparrow$ & 6.1 & 2.3 \\
BD\textsubscript{P}-Rate (\%) $\downarrow$ & -14.1 & -6.2 \\
BD-PSNR-Y (dB) $\uparrow$ & 0.7 & 0.3 \\
BD\textsubscript{S}-Rate (\%) $\downarrow$ & -9.8 & -3.1 \\
BD-SSIM-Yb ($\times$ 100) $\uparrow$ & 0.7 & 0.2 \\
$\Delta$Framedrops (\%) $\downarrow$ & -0.8 & -0.5 \\
\bottomrule
\end{tabular}
\vspace{-.1em}
\end{table}

\begin{table}[ht]
\centering
\caption{Wilcoxon signed-rank test and Cohen's $d$ effect size \cite{cohen2013statistical} of CAE-IGS's improvement vs. Static Ladder across  bitrates and metrics. The top part of the table corresponds to zone 3 of Table \ref{tab: CronosLadder} and the bottom part to zones 0-2. Statistical significance is achieved when $p<0.05$. In such cases, small, medium and large effect sizes correspond to $d < 0.2 $, $0.2 \leq d < 0.8$, $d\geq 0.8$, respectively.}
\label{tab:significance_testing_per_zone}
\begin{tabular}{r | c c | c c | c c}
\toprule
\textbf{Bitrate} & \multicolumn{2}{c|}{\textbf{VMAF}} & \multicolumn{2}{c}{\textbf{SSIM-Yb}} & \multicolumn{2}{c}{\textbf{PSNR-Y}} \\
\textbf{(Mbps)} & $p$ & $d$ & $p$ & $d$ & $p$ & $d$ \\
\midrule
17.5 & 9.44E-01 & -0.08 & 9.99E-01 & -0.02 & 9.99E-01 & -0.09 \\
12.5 & 9.75E-01 & -0.04 & 9.96E-01 & -0.01 & 9.98E-01 & -0.04 \\
\midrule
7.5 & 2.75E-03 & 0.17 & 1.07E-04 & 0.02 & 6.59E-05 & 0.11 \\
3.5 & 9.72E-04 & 0.17 & 5.58E-04 & 0.01 & 1.46E-04 & 0.06 \\
2 & 1.96E-11 & 0.87 & 8.50E-06 & 0.11 & 2.42E-07 & 0.21 \\
\bottomrule
\end{tabular}
\vspace{-1.5em}
\end{table}

Table~\ref{tab:bd_metrics} reports BD metrics and change in frame drop events compared to the Static Ladder. Both CAE-IGS and the Optimal Ladder consistently outperform the Static Ladder across all BD metrics. Notably, both methods also demonstrate an advantage in playback stability, achieving a reduction in frame drop events ($\Delta$Framedrops). CAE-IGS reduces the frame drops by 0.5\% from an average of 3.5\% in the Static Ladder to an average of 3.0\%. Importantly, \textit{only CAE-IGS is practical}: Optimal Ladder assumes that the entire scene is available to encode as a batch and carries out brute-force testing for resolution selection, both of which are incompatible with actual IGS deployments. In contrast, CAE-IGS has inference time of 1ms on a single CPU core and predicts the next-scene resolution using statistics of the current scene, making it straightforward to use in practice.

To validate the statistical significance of the improvement offered by CAE-IGS versus the Static Ladder, we performed the Wilcoxon signed-rank test and calculated Cohen's $d$ effect size \cite{cohen2013statistical} across all target bitrates and video quality metrics. The results are shown in Table~\ref{tab:significance_testing_per_zone}.  
For zones 0-2, the test yields statistically-significant improvement ($p< 0.05$) across all metrics, confirming that CAE-IGS has higher quality that the Static Ladder. For these zones, the effect size is large for VMAF in zone 0 and small for zones 1 and 2. For zone 3, we obtain $ p> 0.05$; i.e., there is not enough evidence to assert visual quality improvement for that zone. This can be empirically observed in the high bitrate points of Fig.~\ref{fig: ExamplesLadders}, where all solutions produce very similar RQ points for zone 3. There, CAE-IGS does offer a modest 3.3\% bitrate reduction vs. the Static Ladder. Future work can explore further resolution adaptation options for this bitrate zone that can push this bitrate saving higher.

\section{Conclusion}
\label{sec: connclusion}
We propose CAE-IGS, a framework for content-adaptive encoding tailored specifically for the conditions of interactive game streaming. Unlike all existing CAE methods that rely on lookahead strategies, full or partial encodings and frame buffering, CAE-IGS introduces per-scene resolution adaptation with no latency overhead. This is achieved through a lightweight CNN model that uses just 1ms of a single CPU core for the entire next-scene inference. Experimentally, CAE-IGS offers higher quality than the Static Ladder, while also reducing bitrate usage and slightly decreasing frame drops. For instance, CAE-IGS achieves -7\% BD-Rate (for VMAF), 2.3 BD-VMAF, and lowers the average frame drop rate from 3.5\% to 3.0\% versus the Static Ladder. While this performance is surpassed by the corresponding offline estimation method (Optimal Ladder), CAE-IGS strikes a good balance between: zero-latency, statistically-significant quality gains in low and medium bitrate, and runtime efficiency.

\bibliographystyle{IEEEtran}
\bibliography{refs}

@IEEEtranBSTCTL{IEEEexample:BSTcontrol,
CTLuse_forced_etal       = "yes",
CTLmax_names_forced_etal = "3",
CTLnames_show_etal       = "2" }

@article{paul2022efficient,
  author={Paul, Somdyuti and Norkin, Andrey and Bovik, Alan C.},
  journal={IEEE Transactions on Image Processing}, 
  title={Convex Hull Prediction for Adaptive Video Streaming by Recurrent Learning}, 
  year={2024},
  volume={33},
  number={},
  pages={5114-5128},
  doi={10.1109/TIP.2024.3455989}}

@inproceedings{wu2020fast,
  title={Fast encoding parameter selection for convex hull video encoding},
  author={Wu, Ping-Hao and Kondratenko, Volodymyr and Katsavounidis, Ioannis},
  booktitle={Applications of digital image processing XLIII},
  volume={11510},
  pages={181--194},
  year={2020},
  organization={SPIE}
}

@book{cohen2013statistical,
  title={Statistical Power Analysis for the Behavioral Sciences},
  author={Cohen, Jacob},
  year={2013},
  publisher={Routledge}
}

@article{bourtsoulatze2019deep,
  title={Deep video precoding},
  author={Bourtsoulatze, Eirina and Chadha, Aaron and Fadeev, Ilya and Giotsas, Vasileios and Andreopoulos, Yiannis},
  journal={IEEE Transactions on Circuits and Systems for Video Technology},
  volume={30},
  number={12},
  pages={4913--4928},
  year={2019},
  publisher={IEEE}
}

@INPROCEEDINGS{HEVC,
    title={{Overview of the high efficiency video coding (HEVC) standard}},
    author = {Sullivan, Gary J. and Ohm, Jens-Rainer and Han, Woo-Jin and Wiegand, Thomas},
    month = dec,
    booktitle={IEEE Transactions on circuits and systems for video technology},
    volume={22},
    number={12},
    pages={1649--1668},
    year={2012},
    publisher={IEEE}
}

@INPROCEEDINGS{menon2024convexhull_xpsnr,
      title={{Convex-hull Estimation using XPSNR for Versatile Video Coding}}, 
      author={{V. V. Menon} and Christian R. Helmrich and Adam Wieckowski and Benjamin Bross and Detlev Marpe},
      booktitle={IEEE International Conference on Image Processing}, 
      year={2024},
}

@INPROCEEDINGS{HLS_ladder_ref,
    author={{Apple Inc.}},
    title = {{HLS Authoring Specification for Apple Devices}},
    url = {https://developer.apple.com/documentation/http_live_streaming/hls_authoring_specification_for_apple_devices}
}

@INPROCEEDINGS{jingwen_vmaf_jnd_sur,
  author={Zhu, Jingwen and Le Callet, Patrick and Perrin, Anne-Flore and Sethuraman, Sriram and Rahul, Kumar},
  booktitle={IEEE International Conference on Image Processing}, 
  title={{On The Benefit of Parameter-Driven Approaches for the Modeling and the Prediction of Satisfied User Ratio for Compressed Video}}, 
  year={2022},
  pages={4213-4217},
  doi={10.1109/ICIP46576.2022.9897946}}

@INPROCEEDINGS{netflix_paper,
  author={De Cock, Jan and Li, Zhi and Manohara, Megha and Aaron, Anne},
  booktitle={2016 IEEE International Conference on Image Processing (ICIP)}, 
  title={{Complexity-based consistent-quality encoding in the cloud}}, 
  year={2016},
  volume={},
  number={},
  pages={1484-1488},
  doi={10.1109/ICIP.2016.7532605}}

@INPROCEEDINGS{VMAF,
  title={{VMAF: The journey continues}},
  author={Zhi Li and Christos Bampis and Julie Novak and Anne Aaron and Kyle Swanson and Anush Moorthy and Jan De Cock},
  booktitle={Netflix Technology Blog},
  volume={25},
  year={2018}
}

@INPROCEEDINGS{ensemble_learning_vvc_ladder,
  author = {Nasiri, Fatemeh and Hamidouche, Wassim and Morin, Luce and Dholland, Nicolas and Aubié, Jean-Yves},
  booktitle={10th European Workshop on Visual Information Processing}, 
  title={{Ensemble Learning for Efficient VVC Bitrate Ladder Prediction}}, 
  year={2022},
  doi={10.1109/EUVIP53989.2022.9922824}}

@inproceedings{HAS_survey,
title = {A {Survey} on {Bitrate} {Adaptation} {Schemes} for {Streaming} {Media} {Over} {HTTP}},
volume = {21},
issn = {1553-877X, 2373-745X},
doi = {10.1109/COMST.2018.2862938},
number = {1},
booktitle = {IEEE Communications Surveys \& Tutorials},
author = {Bentaleb, Abdelhak and Taani, Bayan and Begen, Ali C. and Timmerer, Christian and Zimmermann, Roger},
year = {2019},
pages = {562--585},
}

@ARTICLE{Katsenou_IEEEOJSP2021,
  author={Katsenou, Angeliki and Sole, Joel and Bull, David R.},
  journal={IEEE Open Journal of Signal Processing}, 
  title={Efficient Bitrate Ladder Construction for Content-Optimized Adaptive Video Streaming}, 
  year={2021},
  volume={2},
  pages={496-511},
  doi={10.1109/OJSP.2021.3086691}}

@INPROCEEDINGS{KokaramICIP2018, 
author={Chen, Chao and Lin, Yao-Chung and Benting, Steve and Kokaram, Anil}, 
booktitle={25th IEEE International Conference on Image Processing (ICIP)}, 
title={{Optimized Transcoding for Large Scale Adaptive Streaming Using Playback Statistics}}, 
year={2018}, 
pages={3269-3273}, 
month={Oct},}

@INPROCEEDINGS{Menon_ICME2022,
  author={Menon, Vignesh V and Amirpour, Hadi and Ghanbari, Mohammad and Timmerer, Christian},
  booktitle={IEEE International Conference on Multimedia and Expo}, 
  title={{Perceptually-Aware Per-Title Encoding for Adaptive Video Streaming}}, 
  year={2022},
  doi={10.1109/ICME52920.2022.9859744}}

@INPROCEEDINGS{Krishna_PCS2024,
  author={Durbha, Krishna Srikar and Tmar, Hassene and Stejerean, Cosmin and Katsavounidis, Ioannis and Bovik, Alan C.},
  booktitle={2024 Picture Coding Symposium (PCS)}, 
  title={{Bitrate Ladder Construction Using Visual Information Fidelity}}, 
  year={2024},
  doi={10.1109/PCS60826.2024.10566405}}

@inproceedings{tashtarian2024artemis,
  author    = {Farzad Tashtarian and Abdelhak Bentaleb and Hadi Amirpour and Sergey Gorinsky and Junchen Jiang and Hermann Hellwagner and Christian Timmerer},
  title     = {{ARTEMIS}: Adaptive Bitrate Ladder Optimization for Live Video Streaming},
  booktitle = {USENIX Symposium on Networked Systems Design and Implementation},
  year      = {2024},
  pages     = {591--611},
  }

@ARTICLE{GhasempourJETCAS2025,
  author={Ghasempour, Mohammad and Amirpour, Hadi and Timmerer, Christian},
  journal={IEEE Journal on Emerging and Selected Topics in Circuits and Systems}, 
  title={Real-Time Quality- and Energy-Aware Bitrate Ladder Construction for Live Video Streaming}, 
  year={2025},
  volume={15},
  number={1},
  pages={83-93},
  }

@article{TeliliACMTransMCCA2025,
author = {Telili, Ahmed and Hamidouche, Wassim and Amirpour, Hadi and Fezza, Sid Ahmed and Timmerer, Christian and Morin, Luce},
title = {Convex Hull Prediction Methods for Bitrate Ladder Construction: Design, Evaluation, and Comparison},
year = {2025},
issn = {1551-6857},
journal = {ACM Trans. Multimedia Comput. Commun. Appl.},
month = mar
}

@article{chen2023gamival,
  title={GAMIVAL: video quality prediction on mobile cloud gaming content},
  author={Chen, Yu-Chih and Saha, Avinab and Davis, Chase and Qiu, Bo and Wang, Xiaoming and Gowda, Rahul and Katsavounidis, Ioannis and Bovik, Alan C},
  journal={IEEE Signal Processing Letters},
  volume={30},
  pages={324--328},
  year={2023},
  publisher={IEEE}
}

@ARTICLE{BarmanIEEETCSVT2022,
  author={Barman, Nabajeet and Martini, Maria G.},
  journal={IEEE Transactions on Circuits and Systems for Video Technology}, 
  title={User Generated HDR Gaming Video Streaming: Dataset, Codec Comparison, and Challenges}, 
  year={2022},
  volume={32},
  number={3},
  pages={1236-1249},
  doi={10.1109/TCSVT.2021.3077384}}

@inproceedings{ReznikPV2018,
author = {Reznik, Yuriy A. and Lillevold, Karl O. and Jagannath, Abhijith and Greer, Justin and Corley, Jon},
title = {Optimal Design of Encoding Profiles for ABR Streaming},
year = {2018},
booktitle = {Proceedings of the 23rd Packet Video Workshop},
pages = {43–47},
numpages = {5},
location = {Amsterdam, Netherlands},
series = {PV '18}
}

@INPROCEEDINGS{AfonsoPCS2016,
  author={Afonso, Mariana and Katsenou, Angeliki and Zhang, Fan and Agrafiotis, Dimitris and Bull, David},
  booktitle={2016 Picture Coding Symposium (PCS)}, 
  title={Video texture analysis based on HEVC encoding statistics}, 
  year={2016},
}

@article{katsenou2022study,
  title={Study of compression statistics and prediction of rate-distortion curves for video texture},
  author={Katsenou, Angeliki and Afonso, Mariana and Bull, David R},
  journal={Signal Processing: Image Communication},
  volume={101},
  pages={116551},
  year={2022},
  publisher={Elsevier}
}

@ARTICLE{Wang_SSIM,
  author={Zhou Wang and Bovik, A.C. and Sheikh, H.R. and Simoncelli, E.P.},
  journal={IEEE Transactions on Image Processing}, 
  title={Image quality assessment: from error visibility to structural similarity}, 
  year={2004},
  volume={13},
  number={4},
  pages={600-612},
  doi={10.1109/TIP.2003.819861}}

\end{document}